\documentclass[preprint,3p]{elsarticle}
\usepackage{mymacros}
\usepackage{amssymb}
\usepackage{amsthm}
\usepackage[mathlines]{lineno}
\usepackage{graphicx}
\usepackage{units}
\usepackage{url}
\usepackage{amsmath}
\usepackage{amsfonts}
\usepackage{bm}
\usepackage{textcomp}
\usepackage{subfigure}
\usepackage{multicol}
\usepackage{multirow}
\usepackage{verbatim}
\usepackage{rotating}
\usepackage[colorlinks,linkcolor=red,citecolor=red]{hyperref}
\usepackage{float}
\usepackage{epsfig}
\usepackage{dcolumn}
\usepackage{bm}
\usepackage{color}
\usepackage{pstricks}
\usepackage{pst-node}
\usepackage{times}
\usepackage{indentfirst}
\usepackage[english]{babel}

\journal{Physics Letters B}

\begin{document}

\begin{frontmatter}

\title{{\bf \boldmath Measurements of Absolute  Branching Fractions for $\Lambda^+_c\to\Xi^0K^+$ and $\Xi(1530)^0K^+$}}

\author{
\begin{small}
\begin{center}
M.~Ablikim$^{1}$, M.~N.~Achasov$^{9,d}$, S. ~Ahmed$^{14}$, M.~Albrecht$^{4}$, M.~Alekseev$^{55A,55C}$, A.~Amoroso$^{55A,55C}$, F.~F.~An$^{1}$, Q.~An$^{52,42}$, J.~Z.~Bai$^{1}$, Y.~Bai$^{41}$
% ...
\\
\vspace{0.2cm}
(BESIII Collaboration)\\
\vspace{0.2cm} {\it
$^{1}$ Institute of High Energy Physics, Beijing 100049, People's Republic of China\\
% ...
$^{4}$ Bochum Ruhr-University, D-44780 Bochum, Germany\\
% ...
$^{9}$ G.I. Budker Institute of Nuclear Physics SB RAS (BINP), Novosibirsk 630090, Russia\\
% ...
$^{14}$ Helmholtz Institute Mainz, Johann-Joachim-Becher-Weg 45, D-55099 Mainz, Germany\\
% ...
$^{41}$ Southeast University, Nanjing 211100, People's Republic of China\\
$^{42}$ State Key Laboratory of Particle Detection and Electronics, Beijing 100049, Hefei 230026, People's Republic of China\\
% ...
$^{52}$ University of Science and Technology of China, Hefei 230026, People's Republic of China\\
% ...
$^{55}$ (A)University of Turin, I-10125, Turin, Italy; (B)University of Eastern Piedmont, I-15121, Alessandria, Italy; (C)INFN, I-10125, Turin, Italy\\
% ...
$^{d}$ Also at the Novosibirsk State University, Novosibirsk, 630090, Russia\\
% ...
}\end{center}
\vspace{0.4cm}
\end{small}
}

\begin{abstract}
We report the first measurements of absolute branching fractions for the $W$-exchange-only processes $\Lambda^+_c\to\Xi^0K^+$ and $\Lambda^+_c\to\Xi(1530)^0K^+$ with the double-tag technique, by analyzing an $e^{+}e^{-}$ collision data sample, that corresponds to an integrated luminosity of 567~pb$^{-1}$ collected at a center-of-mass energy of 4.6~GeV by the BESIII detector.
The branching fractions are measured to be $\mathcal{B}(\Lambda^+_c\to\Xi^0K^+)=(5.90\pm0.86\pm0.39)\times10^{-3}$ and $\mathcal{B}(\Lambda^+_c\to\Xi(1530)^0K^+)=(5.02\pm0.99\pm0.31)\times10^{-3}$, where the first uncertainties are statistical and the second systematic.
Our results are more precise than the previous relative measurements.
\\
\\
\text{Keywords:~~$\Lambda_c^+$, $W$ exchange, absolute branching fraction, BESIII}
\end{abstract}
\end{frontmatter}



\begin{multicols}{2}

\section{Introduction}
Weak decays of charmed baryon provide useful information for understanding the interplay of weak and strong interactions, complementary to the information obtained from charmed mesons. 
The lightest charmed baryon $\lambdacp$ is the cornerstone of the whole charmed baryon spectroscopy, and the measurement of the properties of $\lambdacp$ provides essential input for studying heavier charmed baryons, such as singly and doubly charmed baryons~\cite{Geng:2017mxn,Yu:2017zst} and $b$-baryons~\cite{Rosner:2012gj}. 
However, theory development in describing the $\lambdacp$ has been slow~\cite{1992JGKorner, 1994TUppal, 1994PZenczykowsky, 1996LLChau,1997KKSharma, 1998YKohara, Ivanov98, Asner:2008nq}, mostly due to limited understanding of the nontrivial non-factorizable effects involved, especially the $W$-exchange process. 
This is very different from the cases of the $D_{(s)}$ meson decays, where the $W$-exchange amplitude is suppressed by color and helicity symmetries.
Therefore, clean experimental measurements of the $W$-exchange-only process in $\lambdacp$ decays play an important role in the identification of the non-factorizable contribution in different theoretical calculations~\cite{Cheng:2015rra}.

\begin{figure}[H]
\begin{center}
	\includegraphics[width=0.45\linewidth]{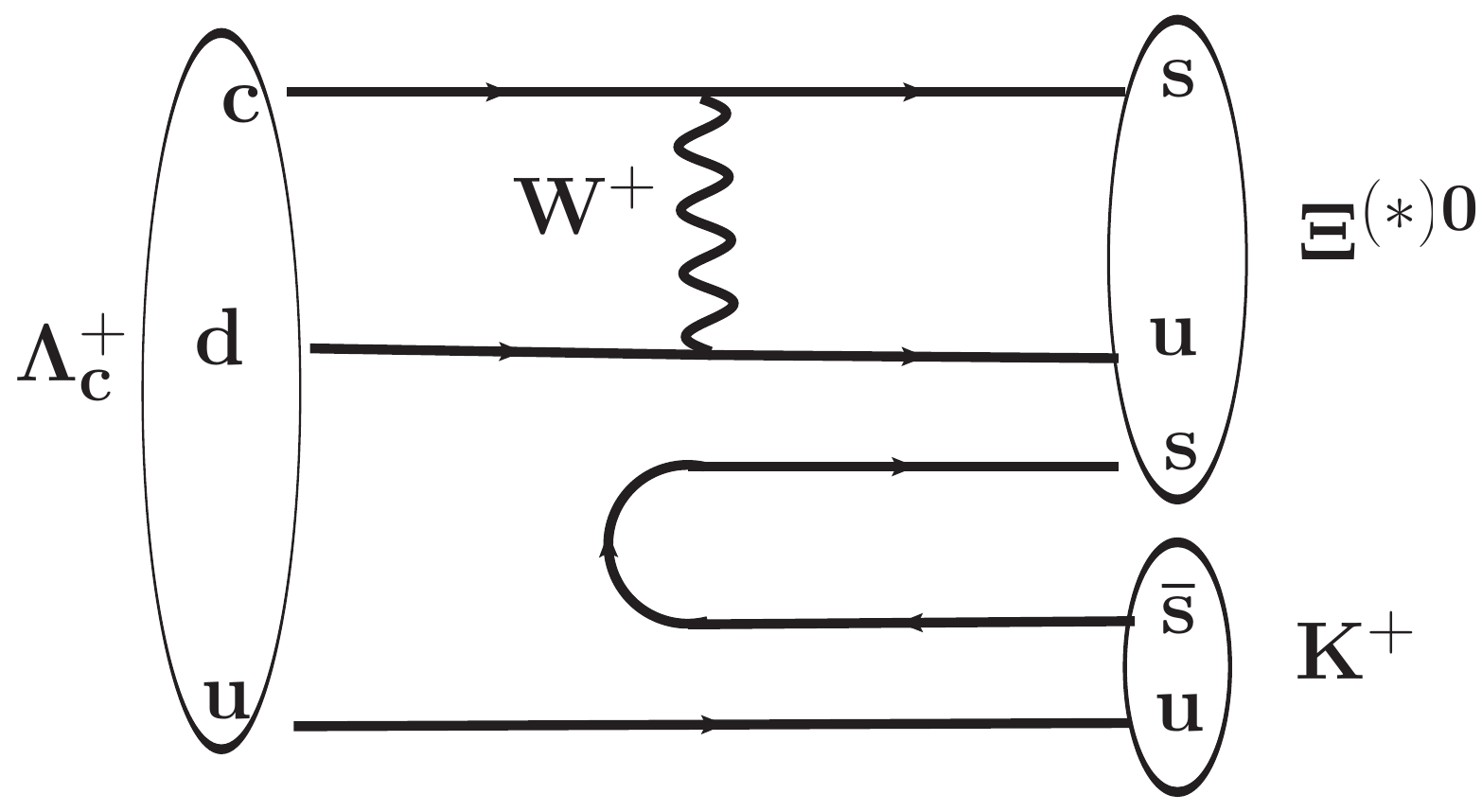}
	\includegraphics[width=0.45\linewidth]{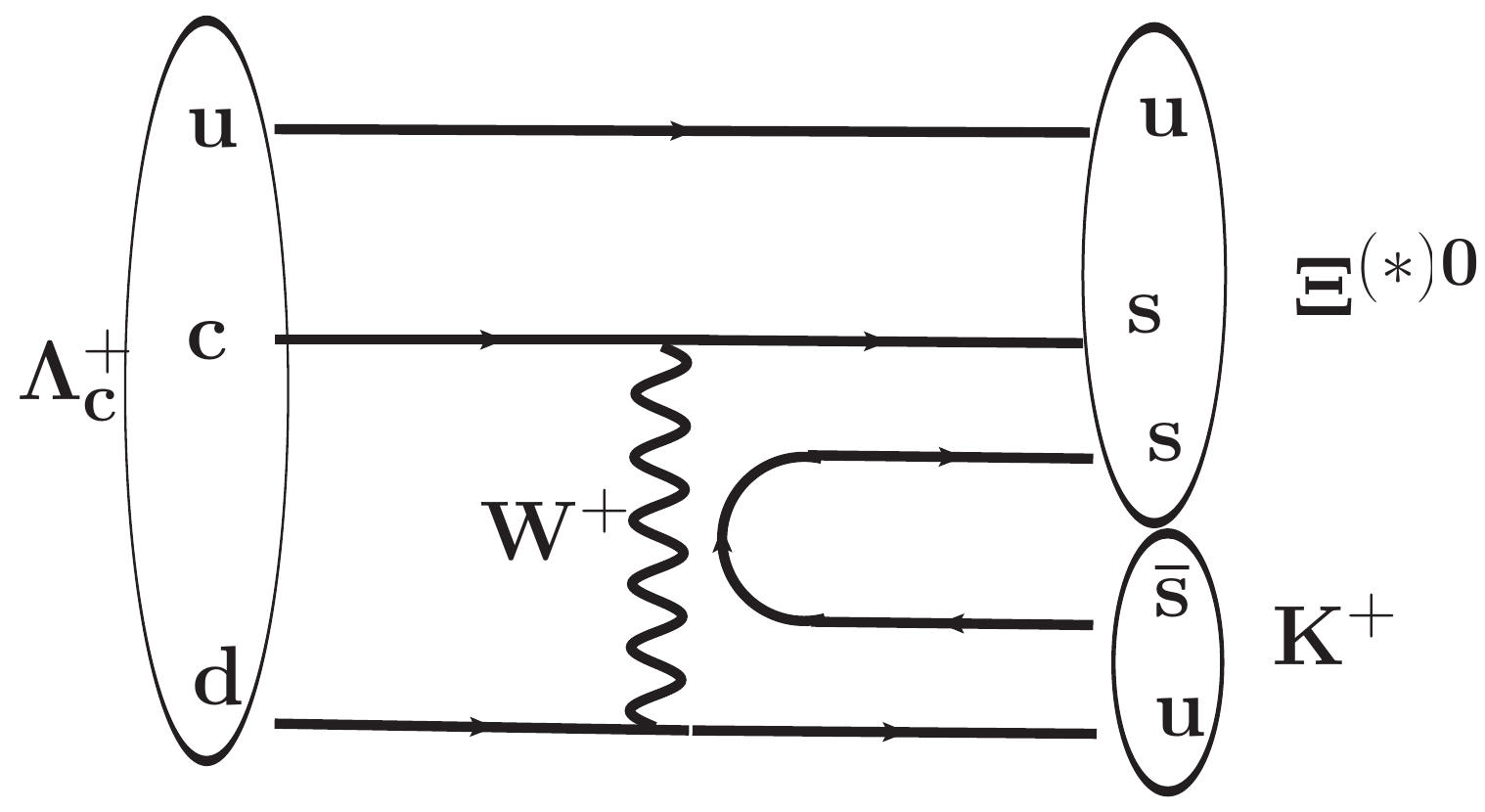}
\caption{ Feynman diagrams of $\LtoXiXisK$.}
\label{fig:feyman}
\end{center}
\end{figure}

The Cabibbo-favored decays $\LtoXiK$ and $\Xi(1530)^0K^+$ proceed only through the $W$-exchange process, as depicted in Fig.~\ref{fig:feyman}. 
These two modes are typical $\lambdacp$ decays to the baryon octet and decuplet states, respectively.
In these two decay modes, large cancellation between different matrix elements occur in both $S$- and $P$-wave decays, making theoretical predictions very difficult~\cite{Cheng:1993gf}.
Several model predictions of the branching fractions ($\mathcal{B}$) for $\Lambda^+_c\to\Xi^{(*)0}K^+$ (here and in the following, $\Xi^{*0}$ is used to denote the $\Xi(1530)^{0}$) are listed in Table~\ref{tab:prediction}. 
They show large variations from each other; the predicted $\mathcal{B}(\LtoXiK)$ fall in the range of $[1.0, 3.6] \times 10^{-3}$~\cite{1992JGKorner,1994PZenczykowsky,Ivanov98, XK92,Verma98}, while the calculations for $\mathcal{B}(\Lambda^+_c\to\Xi^{*0}K^+)$ give three distinct results with one-order-of-magnitude difference~\cite{1992JGKorner,XK92b,Fayyazuddin:1996iy}.
In experiment, these two modes were studied by the CLEO~\cite{Avery:1993vj} and ARGUS~\cite{Albrecht:1994hr} collaborations more than 20 years ago.
Both collaborations directly measured the relative decay rates compared to $\mathcal{B}(\Lmodeb)$, as given in Table~\ref{tab:prediction}. 
Correcting for the branching fraction of the reference channel, $\mathcal{B}(\Lmodeb)$~\cite{belle:CFpkpi, bes:12CF,Amhis:2016xyh}, the average results read $\mathcal{B}(\Lambda^+_c\to\Xi^0K^+)=(5.0\pm1.2)\times10^{-3}$~\cite{pdg2016} and $\mathcal{B}(\Lambda^+_c\to\Xi^{*0}K^+)=(4.0\pm1.0)\times10^{-3}$~\cite{pdg2016,ARGUS:Correct}. 
Apart from the poor precision of the two $\mathcal{B}$'s, the experimental result for $\mathcal{B}(\Lambda^+_c\to\Xi^0K^+)$ exceeds the upper end of the predictions by almost 2$\sigma$.
Hence, an absolute and more precise determination of these $\mathcal{B}$'s is an important input for the modelisation of the hadronic decays of charmed baryons.

\begin{table}[H]
\caption{Comparison of previous experimental measurements and theoretical predictions for $\mathcal{B}(\LtoXiXisK)$.}
\begin{center}
  \resizebox{\linewidth}{!}{
\begin{tabular}{l|c|c|c}
\hline\hline
\multirow{2}*{\minitab[c]{Decay\\ }}                       &  \multirow{2}*{\minitab[c]{Measured $\frac{\mathcal{B}(\Lambda^+_c\to\Xi^{(*)0}K^+)}{\mathcal{B}(\Lambda^+_c\to p K^- \pi^+)}$ \\ }}   & \multirow{2}*{\minitab[c]{Measured \\ $\mathcal{B}(\Lambda^+_c\to\Xi^{(*)0}K^+)$ \\ }}  & \multirow{2}*{\minitab[c]{Predicted \\ $\mathcal{B}(\Lambda^+_c\to\Xi^{(*)0}K^+)$\\ }}  \\
                            &                   &                                 &     \\ \hline
\multirow{5}{*}{$\XiK$}     &  \multirow{5}*{\minitab[c]{$(7.8\pm 1.8)\%$~\cite{Avery:1993vj} \\ }}  &    \multirow{5}*{\minitab[c]{$(5.0\pm1.2)\times10^{-3}$~\cite{pdg2016} \\ }}  &  2.6$\times10^{-3}$~\cite{1992JGKorner}  \\
                            &                   &                                 &  3.6$\times10^{-3}$~\cite{1994PZenczykowsky}  \\
                            &                   &                                 &  3.1$\times10^{-3}$~\cite{Ivanov98}  \\
                            &                   &                                 &  1.0$\times10^{-3}$~\cite{XK92}  \\
                            &                   &                                 &  1.3$\times10^{-3}$~\cite{Verma98}  \\ \hline
\multirow{3}{*}{$\XisK$}    &   \multirow{3}*{\minitab[c]{$(5.3\pm1.9)\%$~\cite{Avery:1993vj}  \\ $(9.3\pm3.2)\%$~\cite{Albrecht:1994hr,ARGUS:Correct} }} & \multirow{3}*{\minitab[c]{$(4.0\pm 1.0)\times10^{-3}$~\cite{pdg2016,ARGUS:Correct} \\ }}  & \multirow{3}*{\minitab[c]{5.0$\times10^{-3}$~\cite{1992JGKorner}  \\  0.8$\times10^{-3}$~\cite{XK92b}  \\  0.6$\times10^{-3}$~\cite{Fayyazuddin:1996iy} }} \\
                            &                   &                                 &     \\ 
                            &                   &                                 &     \\ \hline \hline
\end{tabular}
}
\label{tab:prediction}
\end{center}
\end{table}

In this Letter, we present a study of the $W$-exchange-only decays $\LtoXiK$ and $\Xi(1530)^0K^+$, based on a data sample corresponding to an integrated luminosity of 567\,pb$^{-1}$~\cite{Ablikim:2015nan} collected with the BESIII detector~\cite{Ablikim:2009aa} at the center-of-mass energy of $\sqrt{s}=4.6$\,GeV.
Throughout the text, charge-conjugated modes are implicitly assumed, unless otherwise stated.
At this energy, $\lambdacp$ is always produced in a pair accompanied by a $\lambdacm$, and the remaining phase space does not allow any additional hadrons, hence a double-tag technique~\cite{mark3} can be employed.
This technique does not require a measurement of the luminosity and knowledge of the production cross section, thus providing a model-independent measurement of  $\br{\Lambda_{c}^{+}\to \Xi^{(*)0}K^{+}}$.
First, we select a `single-tag' (ST) sample of $\lambdacm$ candidates by reconstructing the $\lambdacm$ exclusively in one of 12 hadronic decays, as described later.
Then, we search for  $\Lambda_{c}^{+}\to \Xi^{(*)0}K^{+}$ candidates in the system recoiling against the ST side; 
the collection of selected candidates is referred to as the double-tag (DT) sample.
In this analysis, we only detect one $K^{+}$ in the DT side and deduce the presence of a $\Xi^{(*)0}$ in the final state from four-momentum conservation.
The absolute $\mathcal{B}$ of $\Lambda_{c}^{+}\to \Xi^{(*)0}K^{+}$ is then determined from the efficiency-corrected ratio of DT yields to ST yields.


\section{BESIII Detector and Monte Carlo Simulation}
The BESIII detector is a cylindrically symmetric detector with $93\%$ coverage of the full solid angle around the $\ee$ interaction point (IP).
The components of the apparatus are a helium-based main drift chamber (MDC), a plastic scintillator time-of-flight (TOF) system, a 6240-cell CsI(Tl) crystal electromagnetic calorimeter (EMC), a superconducting solenoid providing 1.0\,T magnetic field aligned with the beam axis, and a muon counter with resistive plate chambers as the active element.
The momentum resolution for charged tracks in the MDC is $0.5\%$ at a transverse momentum of $1\gevc$.
The photon energy resolution for $1\gevc$ photos in the EMC is $2.5\%$ in the barrel region and $5.0\%$ in the end-cap region.
The combined information of the ionization energy deposited in the MDC and the flight time measured by the TOF is used to perform particle identification (PID) for charged tracks.
More details about the design and performance of the BESIII detector are given in Ref.~\cite{Ablikim:2009aa}.

We use high-statistics Monte Carlo (MC) simulation samples of $\ee$ annihilations to understand backgrounds and to estimate detection efficiencies.
The $\ee$ annihilation is simulated by the KKMC generator~\cite{Jadach:2000ir}, taking into account the beam energy spread and effects of initial-state radiation (ISR).
The response of the detector to the final-state particles is simulated using {\tt GEANT4}~\cite{ref:geant4}.
Inclusive MC samples, consisting of generic $\lambdacp\lambdacm$ events, $D^{*}_{(s)}\bar{D}^{*}_{(s)}+X$ production~\cite{Brambilla:2010cs}, ISR return to the charmonium(-like) $\psi$ states at lower masses, and continuum processes $\ee\to q\bar{q}~(q=u,d,s)$ are generated to study the backgrounds and to estimate the ST detection efficiencies.
Exclusive DT signal MC events, where the $\lambdacm$ decays into the studied ST modes and the $\lambdacp$ decays into $\XiK$ or $\Xi^{*0}K^+$ (with $\Xi^{0}$ and $\Xi^{*0}$ decaying generically to all known channels), are used to determine the DT detection efficiencies.
All assumed simulated decay rates are taken from in Ref.~\cite{pdg2016}, and the decays are generated using {\tt EVTGEN}~\cite{Lange:2001uf}.

For the MC production of $\LLpair$ events, the observed cross sections are taken into account, and phase space generated $\lambdacm$ decays are re-weighted according to the observed features in data.
For the decays of $\LtoXiXisK$, the angular distributions of $K^{+}$ are generated following $1+\alpha_{\Xi^{(*)}K} \cos^2\theta_K$, where $\theta_K$ is the polar angle of the $K^{+}$ in the rest system of the $\lambdacp$.
The parameters $\alpha_{\Xi^{(*)}K}$ in these two decays are determined from our measurement, as discussed later.

\section{Analysis}
The ST $\lambdacm$ baryon candidates are reconstructed using 12 hadronic decay modes: $\lambdacm \to \Modea$, $\Modeb$, $\Modec$, $\Moded$, $\Modee$, $\Modef$, $\Modeaa$, $\Modebb$, $\Modedd$, $\Modeaaa$, $\Modeccc$, and $\Modeddd$. 
Here, the intermediate particles \mediate are reconstructed through their decays $\Ks \to \pip\pim$, $\bar{\Lambda} \to \bar{p} \pi^+$, $\bar{\Sigma}{}^{0} \to \gamma \bar{\Lambda}$, $\bar{\Sigma}^{-} \to \bar{p} \pi^0$, and $\pi^0 \to \gamma \gamma$, respectively.

Charged tracks are required to satisfy $|\cos\theta| < 0.93$, where $\theta$ is the polar angle with respect to the positron beam direction.
Their distances of closest approaches to the IP are required to be less than 10~cm and 1~cm along and in the plane perpendicular to the electron beam axis, respectively. 
Tracks are identified as protons if their PID likelihood ($\mathcal{L}$) satisfies $\mathcal{L}(p)>\mathcal{L}(K)$ and $\mathcal{L}(p)>\mathcal{L}(\pi)$, while charged kaons and pions are selected using $\mathcal{L}(K)>\mathcal{L}(\pi)$ and $\mathcal{L}(\pi)>\mathcal{L}(K)$, respectively.
More information related to PID in BESIII can be found elsewhere~\cite{Asner:2008nq}.

Clusters in the EMC not associated with any charged track are identified as photon candidates if they satisfy the following requirements:
the deposited energy is required to be larger than 25\,MeV in the barrel region ($|\cos\theta|<0.8$) or 50\,MeV in the end-cap region ($0.86<|\cos\theta|<0.92$).
To suppress background from electronic noise and showers unrelated to the event, the shower time measured by the EMC relative to the event start time is required to be between 0 and 700 ns.
The $\pi^{0}$ candidates are reconstructed from photon pairs that have an invariant mass satisfying $115<M(\gamma\gamma)<150\mevcc$.
To improve the momentum resolution, a kinematic fit constraining the invariant mass to the $\pi^{0}$ nominal mass~\cite{pdg2016} is applied to the photon pairs and the resulting energy and momentum of the $\pi^0$ are used for further analysis.

Candidates for $\Ks$ and $\bar{\Lambda}$ are formed by combining two oppositely charged tracks of $\pip\pim$ and $\bar{p}\pip$, respectively.
For these two tracks, their distances of closest approaches to the IP must be within $\pm$20\,cm along the electron beam direction.
No distance constraints in the transverse plane are required.
The two daughter tracks are constrained to originate from a common decay vertex by requiring the $\chi^2$ of the vertex fit to be less than 100.
Furthermore, the decay vertex is required to be separated from the IP by a distance greater than twice the fitted vertex resolution.
In this procedure, as the combinational backgrounds have been highly suppressed, the charged pions are not subjected to the PID requirement described above, to have the optimal signal significance.
The vertex fitted momenta of the daughter particles are used in the further analysis.
We impose the requirements $487<M(\pi^+\pi^{-})<511\,\mevcc$ and $1111<M(\bar{p}\pi^{+})<1121\,\mevcc$ for $\Ks$ and $\bar{\Lambda}$ candidates.
The $\bar{\Sigma}{}^{0}$ and $\bar{\Sigma}{}^{-}$ candidates are reconstructed from any combinations of $\bar{\Lambda}\gm$ and $\bar{p}\pizero$ with requirement $1179<M(\bar{\Lambda}\gm)<1203\mevcc$ and $1176<M(\bar{p}\pizero)<1200\mevcc$, respectively.
The above requirements on the invariant masses correspond to approximately $\pm 3$ standard deviations around the nominal masses~\cite{pdg2016}.
For the decay modes $\Modec$, $\Moded$, $\Modef$ and $\Modeddd$, possible backgrounds including $\bar{\Lambda}\to \bar{p} \pip$ in the final state are rejected by requiring $M(\bar{p}\pip)$ to be out of the range $(1110, 1120)\mevcc$.
In addition, $\Modec$ candidates satisfying $1170<M(\bar{p}\pi^0)<1200\mevcc$ are excluded to suppress the backgrounds with a $\bar{\Sigma}{}^{-}$ in the final state.
To remove $\Ks$ candidates in the modes $\Modef$, $\Modedd$, $\Modeccc$ and $\Modeddd$, the mass of any $\pip\pim$ and $\pi^0\pi^0$ pair is not allowed to fall in the range (480, 520)$\mevcc$.

\begin{figure}[H]
\centering
\includegraphics[width=\linewidth]{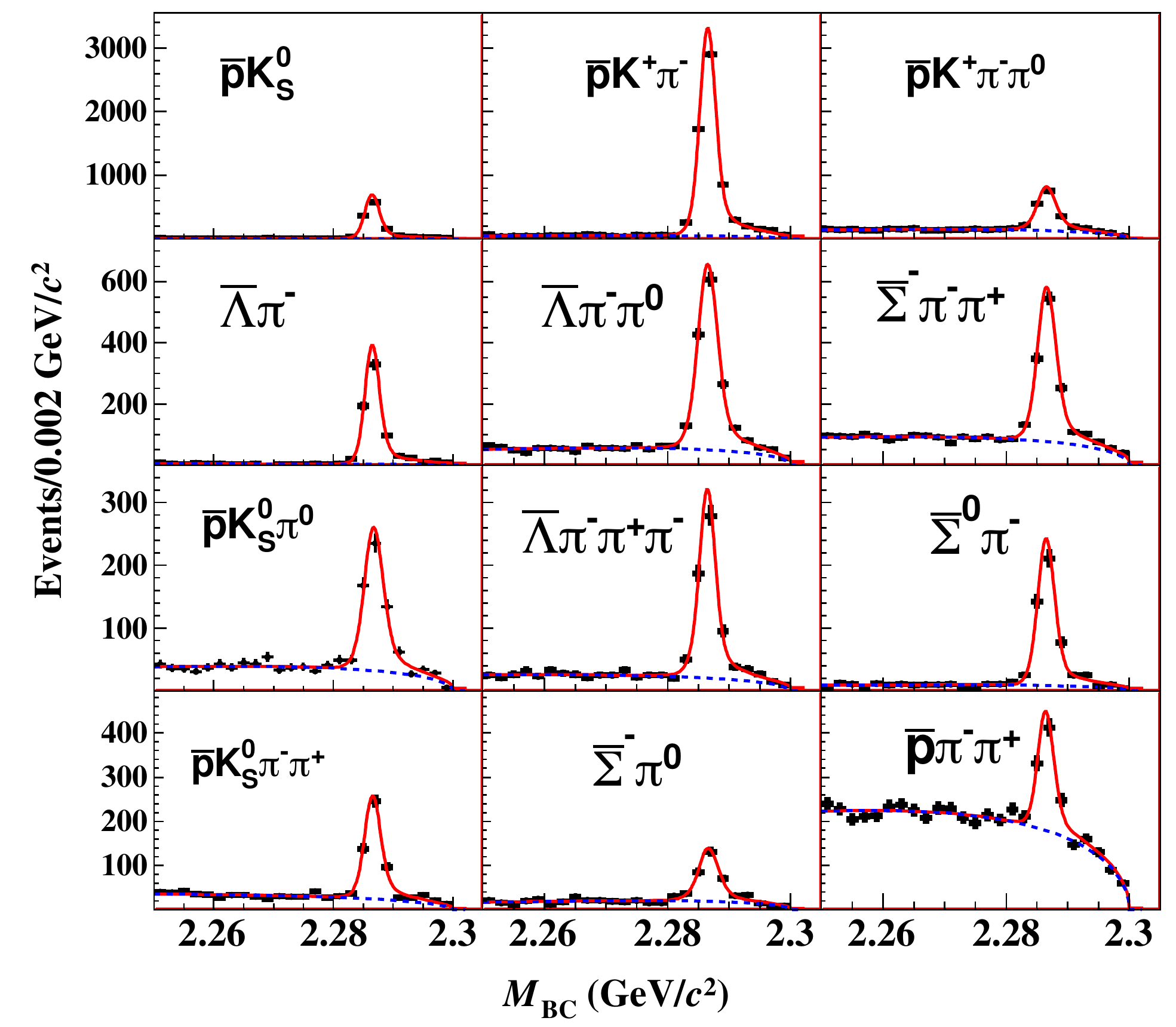}
\caption{Fits to the $M_{\rm BC}$ distributions in data for the different ST modes. Points with error bars are data, solid lines are the sum of the fit functions, and dashed lines are the background shapes.}
\label{fig:ST_datafit}
\end{figure}

The ST $\lambdacm$ yields are identified using the beam-constrained mass $M_{\rm BC}\equiv \sqrt{E_{\rm beam}^2/c^4-p^2/c^2}$, where  $E_{\rm beam}$ is the average value of the $e^+$ and $e^-$ beam energies and $p$ is the measured $\lambdacm$ momentum in the center-of-mass system of the $\ee$ collision.
To improve the signal purity, the energy difference $\Delta{}E \equiv E - E_{\rm beam}$ for the $\lambdacm$ candidate is required to fulfill a mode-dependent $\Delta{}E$ requirement shown in Table~\ref{tab:STyields}, corresponding to approximately three times the resolutions.
Here, $E$ is the total reconstructed energy of the $\lambdacm$ candidate.
For each ST decay mode, if more than one candidate satisfies the above requirements, we select the one with minimal $|\Delta{}E|$.
Figure~\ref{fig:ST_datafit} shows the $M_{\rm BC}$ distributions for the ST samples, where evident $\lambdacm$ signals peak at the nominal $\lambdacm$ mass~\cite{pdg2016}.
We follow the procedure described in Ref.~\cite{bes:12CF} to determine the ST yields for a given tag mode $i$ [$N_{i}^{\rm ST}$] in the signal region $2282<M_{\rm BC}<2291\mevcc$ and the corresponding detection efficiencies [$\varepsilon_{i}^{\rm ST}$], as summarized in Table~\ref{tab:STyields}.
In the procedure of extracting detection efficiencies, the $M_{\rm BC}$ resolutions in MC samples are corrected to agree with those in data.
Besides, MC simulations show that peaking backgrounds in some ST modes are observed with a yield of $1\%$ or less that of the signal.

\begin{table}[H]
  \begin{center}
 \caption{Requirements on $\Delta{}E$, ST yields $N_{i}^{\rm ST}$ and detection efficiencies $\varepsilon_{i}^{\rm ST}$, and DT efficiencies of $\varepsilon_{i,\Xi K}^{\rm DT}$ and $\varepsilon_{i,\Xi^{*}K}^{\rm DT}$. The uncertainties are statistical only. The quoted efficiencies do not include any subleading $\mathcal{B}$.}
  \resizebox{\linewidth}{!}{
  \begin{tabular}{l|c|c|c|c|c}
      \hline \hline
  Mode  &$\Delta{}E$ (MeV) & $N_{i}^{\rm ST}$ & $\varepsilon_{i}^{\rm ST}(\%)$ & $\varepsilon_{i,\Xi K}^{\rm DT}(\%)$  & $\varepsilon_{i,\Xi^{*}K}^{\rm DT}(\%)$\\ \hline
$\textbf{$\Modea$}$ & $(-20,20)$ & $1145\pm34$ & $51.6$ & $41.2$ & $42.6$  \\
$\textbf{$\Modeb$}$ & $(-20,20)$ & $5722\pm80$ & $45.2$ & $37.3$ & $39.1$ \\
$\textbf{$\Modec$}$ & $(-30,20)$ & $478\pm28$ & $17.2$ & $15.1$ & $15.2$ \\
$\textbf{$\Moded$}$ & $(-20,20)$ & $431\pm25$ & $18.6$ & $15.4$ & $15.2$ \\
$\textbf{$\Modee$}$ & $(-30,20)$ & $1407\pm51$ & $14.7$ & $13.4$ & $12.7$ \\
$\textbf{$\Modef$}$ & $(-20,20)$ & $474\pm41$ & $55.4$ & $43.3$ & $45.1$ \\
$\textbf{$\Modeaa$}$ & $(-20,20)$ & $648\pm25$ & $38.7$ & $30.9$ & $31.4$ \\
$\textbf{$\Modebb$}$ & $(-30,20)$ & $1282\pm43$ & $13.0$ & $10.9$ & $11.2$ \\
$\textbf{$\Modedd$}$ & $(-20,20)$ & $540\pm27$ & $10.6$ & $9.0$ & $8.8$ \\
$\textbf{$\Modeaaa$}$ & $(-20,20)$ & $427\pm23$ & $24.1$ & $20.6$ & $20.6$ \\
$\textbf{$\Modeccc$}$ & $(-50,30)$ & $258\pm20$ & $19.6$ & $17.3$ & $17.4$ \\
$\textbf{$\Modeddd$}$ & $(-30,20)$ & $1005\pm42$ & $20.1$ & $17.2$ & $18.1$ \\ 
\hline  \hline
   \end{tabular}
   }
   \label{tab:STyields}
  \end{center}
  \end{table}

Candidates of $\LtoXiXisK$ decays are reconstructed from the remaining tracks recoiling against the ST $\lambdacm$.
A kaon with opposite charge to the tagged $\lambdacm$ is selected with the same selection criteria as described above.
No multiple DT candidates in an event are observed.
The kinematic variable 
\begin{eqnarray}
M_{\rm miss} &\equiv& \sqrt{E_{\rm miss}^{2}/c^4-|\vec{{p}}_{\rm miss}|^{2}/c^2},
\end{eqnarray}
is used to infer the undetected $\Xi^{0}$ and $\Xi^{*0}$,
where $E_{\rm miss}$ and $\vec{{p}}_{\rm miss}$ are the missing energy and momentum carried away by the undetected $\Xi^{0}$ or $\Xi^{*0}$.
The $E_{\rm miss}$ and $\vec{{p}}_{\rm miss}$ are calculated by $E_{\rm miss} \equiv E_{\rm beam}-E_{\it K^{+}}$ and $\vec{{p}}_{\rm miss} \equiv \vec{{p}}_{\lambdacp} - \vec{{p}}_{\it K^{+}}$, where $E_{\it K^{+}}$($\vec{{p}}_{\it K^{+}}$) is the energy (momentum) of the $K^{+}$ in the $\ee$ center-of-mass system.
The momentum of the $\lambdacp$ baryon $\vec{{p}}_{\lambdacp}$ is calculated by $\vec{{p}}_{\lambdacp} \equiv -\hat{p}_{\rm tag} \sqrt{E_{\rm beam}^{2}/c^2-m_{\lambdacp}^{2} c^2}$, where $\hat{p}_{\rm tag}$ is the momentum direction of the ST $\lambdacm$ and $m_{\lambdacp}$ is the nominal mass of the $\lambdacp$~\cite{pdg2016}.
For the signal $\Lambda_{c}^{+}\to \Xi^{(*)0}K^{+}$ decay, $M_{\rm miss}$ is expected to peak at the the nominal masses of the $\Xi^{0}$ and $\Xi^{*0}$, \emph{i.e.} at $1314.9\mevcc$ and $1531.8\mevcc$, respectively~\cite{pdg2016}.

\begin{figure}[H]
\centering
\includegraphics[width=\linewidth]{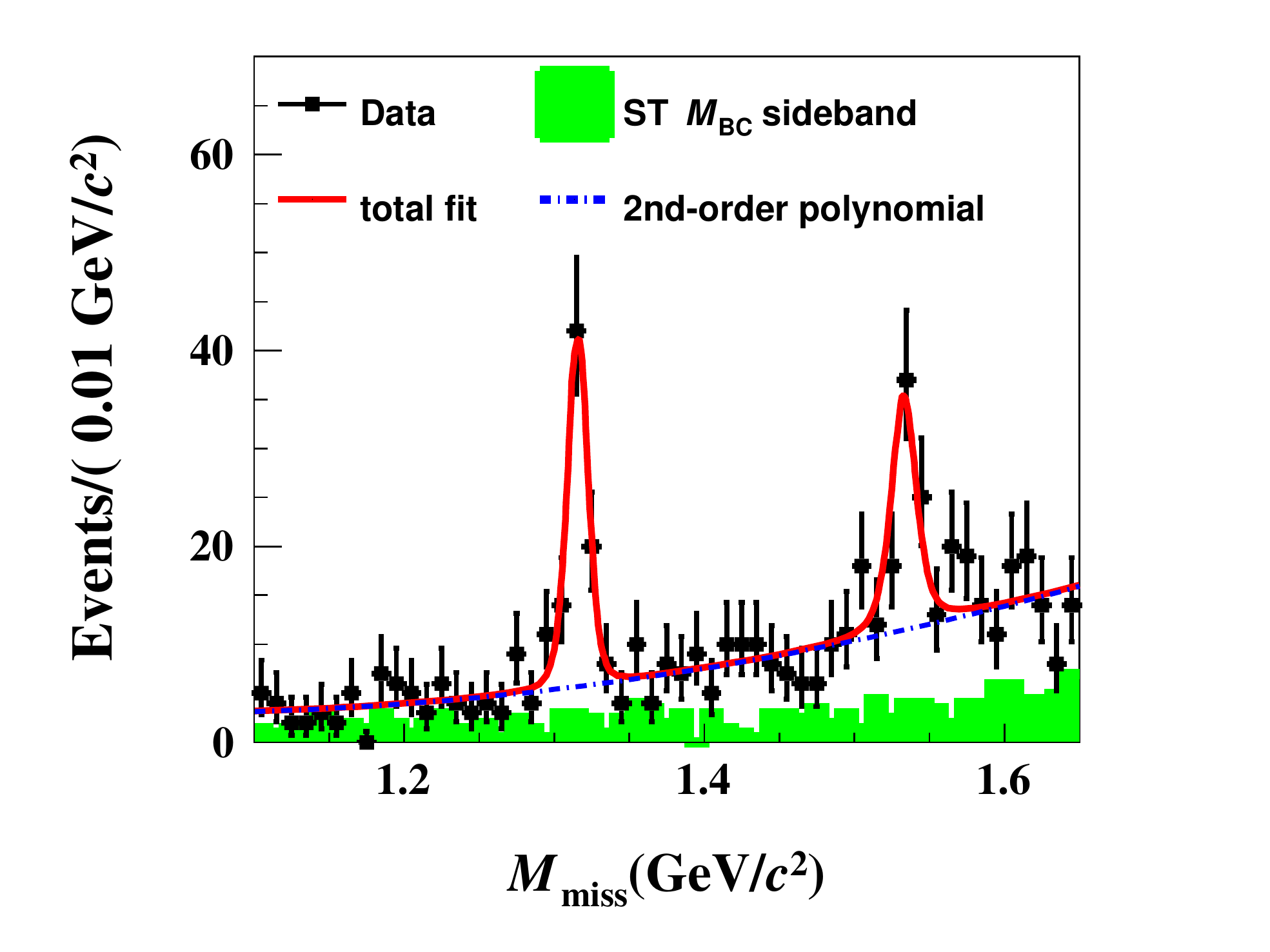}
\caption{Fit to the $M_{\rm miss}$ distribution of the DT candidates. Points with error bars are data, the solid line is the sum of fit functions, the dash-dotted line is the quadratic function for background and the shaded histogram shows normalized data from the ST $M_{\rm BC}$ sideband region, defined as $2.250<M_{\rm BC}<2.265\,\gevcc$.}
\label{fig:DT_datafit}
\end{figure}

We combine the DT candidates over the 12 ST modes and plot the resulting $M_{\rm miss}$ distribution in Fig.~\ref{fig:DT_datafit}.
An unbinned maximum likelihood fit is performed to determine DT signal yields.
The $\Lambda_{c}^{+}\to \Xi^{(*)0}K^{+}$ signal shape is obtained from the MC-derived signal shape convolved with a Gaussian function common to both signal channels whose parameters are left free in the fit.
The background shape is described by a quadratic function, which is validated by the candidate events in the ST $M_{\rm BC}$ sideband region of data and the MC-simulated background samples.
Figure~\ref{fig:DT_datafit} shows the fitted curves to the $M_{\rm miss}$ distribution.
We obtain the DT signal yields of $\XiK$ and $\Xi^{*0}K^+$ to be $N^{\rm DT}_{\Xi K}=68.2\pm9.9$ and $N^{\rm DT}_{\Xi^{*}K}=59.5\pm11.7$, respectively, where the uncertainties are statistical only. 
The statistical significances for the signal are evaluated by the changes in the likelihood between the nominal fit and a fit with the signal yield set to zero;
they are $10.3\sigma$ for $\Xi^0K^+$ and $6.4\sigma$ for $\Xi^{*0}K^+$.

The absolute $\mathcal{B}$ for $\LtoXiK$ and $\LtoXisK$ are obtained by the following formula
\begin{eqnarray}
\mathcal{B}(\Lambda_{c}^{+}\to \Xi^{(*)0}K^{+}) &=& \frac{N_{\Xi^{(*)}K}^{\rm DT}}{\sum_{i}\frac{N_{i}^{\rm ST}}{\varepsilon_{i}^{\rm ST}}\varepsilon_{i,\Xi^{(*)}K}^{\rm DT}}.
\label{eq:bs}
\end{eqnarray} 
The DT efficiencies $\varepsilon_{i,\Xi^{(*)}K}^{\rm DT}$ are evaluated based on the yields of the DT signal MC samples in the $M_{\rm miss}$ signal window $(1.10, 1.65)\gevcc$, as summarized in Table~\ref{tab:STyields}.
Using Eq.~\eqref{eq:bs}, we obtain $\mathcal{B}(\LtoXiK)=(5.90\pm0.86\pm0.39)\times10^{-3}$ and $\mathcal{B}(\LtoXisK)=(5.02\pm0.99\pm0.31)\times10^{-3}$, where the first uncertainties are statistical and the second systematic as described below.

As the DT technique is adopted, the systematic uncertainties originating from reconstructing the ST side cancel.
The systematic uncertainties in the $\mathcal{B}(\LtoXiK)$ and $\mathcal{B}(\LtoXisK)$ measurements mainly arise from possible differences between the data and MC simulation of signal processes, $K^{+}$ tracking, $K^{+}$ PID, the fit to the $M_{\rm miss}$ distribution, ST peaking backgrounds, and the $M_{\rm BC}$ ST distributions. 
The detailed estimation of the different systematic uncertainties are given below.

The signal processes $\Lambda^+_c\to\Xi^{(*)0}K^+$ are simulated by taking into account the angular dependences $1+\alpha_{\Xi^{(*)}K} \cos^2\theta_K$. 
We obtain the parameters $\alpha_{\Xi K}=0.77\pm0.78$ and $\alpha_{\Xi^*K}=-1.00\pm0.34$ from fits to data, where the statistical uncertainties are dominant. 
The uncertainties from signal MC modeling are determined to be 3.2$\%$ and 3.9$\%$ for $\LtoXiK$ and $\LtoXisK$ respectively, by changing the parameter $\alpha_{\Xi^{(*)}K}$ within the uncertainties.

The uncertainties associated with $K^+$ tracking and PID are estimated to be 1.0\% each by studying a set of control samples of $\ee\to K^+K^-\pi^+\pi^-$ events selected from data taken at energies above $\sqrt{s}=4.0$\,GeV.
The uncertainties due to the fit procedure are estimated to be 5.2$\%$ and 3.7$\%$ for $\LtoXiK$ and $\LtoXisK$, respectively, by varying the fit range and background shape.
In order to estimate the overall uncertainties due to the ST peaking backgrounds, we estimate the ratio of peaking background contributing to the total ST yields for each ST mode, and then reweight these ratios by the ST yields $N_{i}^{\rm ST}$ obtained in data.
We evaluate the resultant systematic uncertainties to be 0.8\% for both $\LtoXiK$ and $\LtoXisK$.
A possible bias to the efficiency ratio of the DT and ST selections due to the $M_{\rm BC}$ resolution correction is explored by removing the corresponding correction in MC samples.
The efficiency ratios are re-calculated and the deviations of 2.2$\%$ and 2.4$\%$ to the nominal results are taken as the systematic uncertainties for $\LtoXiK$ and $\LtoXisK$, respectively.
All these systematic uncertainties are summarized in Table~\ref{tab:sys_err}, and the total systematic uncertainties are evaluated to be 6.7\% and 6.1\% for $\LtoXiK$ and $\LtoXisK$, respectively, by summing up all the contributions in quadrature.
\begin{table}[H]
  \begin{center}
  \caption{Sources of systematic uncertainties and the corresponding relative values.}
  \begin{tabular}{l|c|c}
      \hline \hline
			Source            &  $\XiK(\%)$ &  $\Xi^{*0}K^+(\%)$ \\
			\hline  
			MC model          &    3.2     &     3.9     \\
			Tracking          &    1.0     &     1.0     \\
			PID               &    1.0     &     1.0     \\
			Fitting           &    5.2     &     3.7     \\
      ST peaking background   &    0.8     &     0.8     \\
			$M_{\rm BC}$ requirement      &    2.2     &     2.4    \\
			\hline
      Total             &    6.7      &    6.1     \\
      \hline\hline
    \end{tabular}
    \label{tab:sys_err}
  \end{center}
  \end{table}

\section{Summary}
To summarize, the absolute branching fractions of two $W$-exchange-only processes $\LtoXiK$ and $\Lambda_{c}^{+}\to \Xi{(1530)}^{0}K^{+}$ are measured by employing a double-tag technique, based on a sample of threshold produced data at $\sqrt{s}=4.6\gev$ collected with BESIII detector.
The results are $\mathcal{B}(\Lambda^+_c\to\Xi^0K^+)=(5.90\pm0.86\pm0.39)\times10^{-3}$ and $\mathcal{B}(\Lambda^+_c\to\Xi(1530)^0K^+)=(5.02\pm0.99\pm0.31)\times10^{-3}$, where the first uncertainties are statistical and the second systematic.
These are the first absolute  measurements of the branching fractions for the $\Lambda^+_c\to\Xi^0K^+$ and $\Lambda^+_c\to\Xi(1530)^0K^+$ decays.
The results are consistent with the previous measurements~\cite{Avery:1993vj,Albrecht:1994hr}, but have improved precision.
For the $\XiK$ mode, the combined $\mathcal{B}$ gives $(5.56\pm0.74)\times10^{-3}$, which shows more significant deviations from predicted values in Table~\ref{tab:prediction} by at least 2.6$\sigma$. 
The measured $\mathcal{B}(\Lambda^+_c\to\Xi(1530)^0K^+)$ favors the calculation in Ref.~\cite{1992JGKorner}, while $\mathcal{B}(\Lambda^+_c\to\Xi^0K^+)$ in Ref.~\cite{1992JGKorner} has 4$\sigma$ discrepancy from experimental result.
This indicates that our results are essential to calibrate the $W$-exchange diagram amplitudes in these theoretical approaches.

\section{Acknowledgments}
The BESIII collaboration thanks the staff of BEPCII and the IHEP computing center for their strong support. 
This work is supported in part by National Key Basic Research Program of China under Contract No. 2015CB856700;
National Natural Science Foundation of China (NSFC) under Contracts Nos. 11235011, 11335008, 11425524, 11625523, 11635010;
the Chinese Academy of Sciences (CAS) Large-Scale Scientific Facility Program;
the CAS Center for Excellence in Particle Physics (CCEPP);
Joint Large-Scale Scientific Facility Funds of the NSFC and CAS under Contracts Nos. U1332201, U1532257, U1532258;
CAS Key Research Program of Frontier Sciences under Contracts Nos. QYZDJ-SSW-SLH003, QYZDJ-SSW-SLH040;
CAS under Contracts Nos. KJCX2-YW-N29, KJCX2-YW-N45, QYZDJ-SSW-SLH003;
100 Talents Program of CAS;
National 1000 Talents Program of China; 
INPAC and Shanghai Key Laboratory for Particle Physics and Cosmology;
German Research Foundation DFG under Contracts Nos. Collaborative Research Center CRC 1044, FOR 2359;
Istituto Nazionale di Fisica Nucleare, Italy;
Koninklijke Nederlandse Akademie van Wetenschappen (KNAW) under Contract No. 530-4CDP03;
Ministry of Development of Turkey under Contract No. DPT2006K-120470;
National Science and Technology fund;
The Swedish Research Council;
U. S. Department of Energy under Contracts Nos. DE-FG02-05ER41374, DE-SC-0010118, DE-SC-0010504, DE-SC-0012069;
University of Groningen (RuG) and the Helmholtzzentrum fuer Schwerionenforschung GmbH (GSI), Darmstadt;
WCU Program of National Research Foundation of Korea under Contract No. R32-2008-000-10155-0;
China Postdoctoral Science Foundation;
This paper is also supported by Beijing municipal government under Contract No. CIT\&TCD201704047. 


\end{multicols}
\end{document}